\documentclass[10pt,aps,amsmath,showpacs,amssymb,nofootinbib,superscriptaddress,prl,twocolumn]{revtex4-1}

\usepackage{amssymb}
\usepackage{amsfonts}
\usepackage{rotating}
\usepackage{amsmath}
\usepackage{palatino}
\usepackage{epsfig}
\usepackage{hyperref}

\begin{document}

\title{Random walks and search in time-varying networks}

\author{Nicola Perra}

\affiliation{Laboratory for the Modeling of Biological and
  Socio-technical Systems, Northeastern University, Boston MA 02115
  USA}

\author{Andrea Baronchelli}

\affiliation{Laboratory for the Modeling of Biological and
  Socio-technical Systems, Northeastern University, Boston MA 02115
  USA}

\author{Delia Mocanu}

\affiliation{Laboratory for the Modeling of Biological and
  Socio-technical Systems, Northeastern University, Boston MA 02115
  USA}

\author{Bruno~Gon\c calves}

\affiliation{Laboratory for the Modeling of Biological and
  Socio-technical Systems, Northeastern University, Boston MA 02115
  USA}

\author{Romualdo Pastor-Satorras}

\affiliation{Departament de F\'{\i}sica i Enginyeria Nuclear,
  Universitat Polit\`ecnica de Catalunya, Campus Nord B4, 08034
  Barcelona, Spain}

\author{Alessandro Vespignani}

\affiliation{Laboratory for the Modeling of Biological and
  Socio-technical Systems, Northeastern University, Boston MA 02115
  USA} 

\affiliation{Institute for Scientific Interchange Foundation, Turin
  10133, Italy}

\date{\today} 

\begin{abstract} 
  The random walk process underlies the description of a large
  number of real world phenomena. Here we provide 
  the study of random walk processes in time varying networks in
  the regime of time-scale mixing; i.e. when the network connectivity
  pattern and the random walk process dynamics are unfolding on the
  same time scale. We consider a model for time varying networks
  created from the activity potential of the nodes, and derive
  solutions of the asymptotic behavior of random walks and the mean
  first passage time in undirected and directed networks. Our findings
  show striking differences with respect to the well known results
  obtained in quenched and annealed networks, emphasizing the effects
  of dynamical connectivity patterns in the definition of proper
  strategies for search, retrieval and diffusion processes in
  time-varying networks.
\end{abstract}

\maketitle

Random walks on networks lie at the core of many real-world dynamical
processes, ranging from the navigation and ranking of information
networks to the spreading of diseases and the routing of information
packets in large-scale infrastructures such as the Internet
\cite{barrat08-1,havlin-book,newman10-1,albert02,boccaletti06-1,alex12-1}. In
recent years, empirical evidence pointing out the heterogeneous
topology of many real-world networks has led to a large body of work
focusing on the properties of random walks in networks characterized
by heavy tailed degree distributions and other features such as
clustering and community
structure~\cite{barrat08-1,kleinberg00-1,noh04,gallos04-1,kozma05-1,samukhin08-1,PhysRevE.78.011114}. Although
these studies provided a deeper understanding of processes of
technological relevance such as \emph{WWW} navigation and ranking, they have
mostly focused on the situation in which the time scale characterizing
the changes in the structure of the network and the time scale
describing the evolution of the process are well
separated~\cite{brin98-1,kleinberg98-1,adamic01-1,watts02-1,kleinberg06-1}. While
convenient for analytical tractability, this limit is far from
realistic in many systems including modern information networks, the diffusion and search of information in microblogging systems and social networking platforms, sexually transmitted diseases or the diffusion of ideas and knowledge
in social contexts. In all these cases, the concurrence of contacts
and their dynamical patterns are typically characterized by a time
scale comparable to that of the diffusion
process, motivating the study models able to account for effects of the time varying nature of networks on dynamical processes~\cite{morris93-1,volz09-1,schwartz10-2,holme11-1,jolad11-1,perra12-1,starnini_rw_temp_nets,basu,toro_2007,butts_2008} .

Motivated by the above  problems we study the random walk process 
in a fairly general class of time-varying networks. 
Namely, we consider the {\it activity driven} class of models for time-varying
networks presented in Ref.~\cite{perra12-1} that allows for an
explicit representation of dynamical connectivity patterns. We derive the analytical solutions of the stationary state of the
random walk and the mean first passage time \cite{Redner01} in both
directed and undirected time varying networks. We find that the
behavior of the random walk and the ensuing network discovery process
in time varying networks is strikingly different from those occurring
in quenched and annealed topologies~\cite{noh04,barrat08-1,newman10-1,
  baronchelli2010mean}. These results have the potential to become a
starting point for the definition of alternative strategies and
mechanisms to explore and retrieve information from networks, and
characterize more accurately spreading and diffusion processes in a
wide range of dynamic social networks.

Activity driven network models focus on the activity pattern of each
node, which is used to explicitly model the evolution of the
connectivity pattern over time. Each node $i$ is characterized by a quenched, fixed,
activity rate $a_i$, extracted from a distribution $F\left( a\right)$,
that represents the probability per unit time that a given node will
engage in an interaction and generate the corresponding edge
connecting it with other nodes in the system. In the simplest
formulation of the model, networks are generated according to the
following memoryless stochastic process \cite{perra12-1}: i) At each
time step $t$, the instantaneous network $G_t$ starts with $N$ disconnected nodes;
ii) With probability $a_i \Delta t$, each vertex $i$ becomes active
and generates $m$ undirected links that are connected to $m$ other
randomly selected vertices. Non-active nodes can still receive
connections from other active vertices; iii) At time $t + \Delta t$,
all the edges in the network $G_t$ are deleted and the process starts
over again to generate the network $G_{t+\Delta t}$.  It can be shown
that the full dynamics of the network is encoded in the activity rate
distribution, $F\left( a\right)$, and that the time-aggregated
measurement of network connectivity yields a degree distribution that
follows the same functional form as the distribution $F(a)$.  This
distribution can be assumed a priori or derived from empirical data in
the case of high quality data set such as social/information
networks~\cite{perra12-1}.

\begin{figure}
  \begin{center}
    \includegraphics [width=0.5\textwidth] {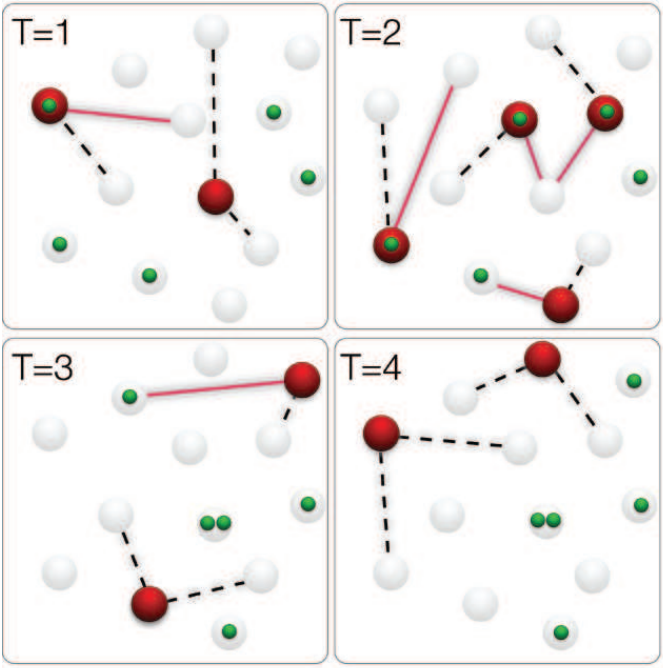}
    \caption{Activity driven random walk process. Active nodes are
      shown in red and walkers are presented in green. Links used by
      walkers to move from one node to another are shown in solid red
      lines, while edges connecting empty nodes are shown as dashed
      lines. Here we considered $m=2$.}
    \label{rw}
  \end{center}
\end{figure}

Although the above model is memoryless, it can be considered as the simplest yet non trivial setting for the study of the concurrence of changes in connectivity pattern of the network and the dynamical process unfolding on its structure. 
We define the random walk process on time varying networks as follows:
at each time step $t$ the network $G_t$ is generated, and the walker
diffuses for a time $\Delta t$. After diffusion, at time $t+\Delta t$,
a new network $G_{t+\Delta t}$ is generated (see Figure~\ref{rw}). The
concurrent dynamics of the random walker and the network take thus
place with the same time scale, which introduces a feature not found
in the equivalent processes in quenched or annealed networks, namely
that walkers can get trapped in temporarily isolated nodes. 
In other words, the diffusive dynamic of the particles is "enslaved" to the local connectivity pattern of each node so that effectively the diffusive process is transformed in a "transport" process defined concurrently by the network dynamic and the particle diffusive process. 

The probability $P_i(t)$ that a random walker is in node $i$ at time
$t$ obeys the  master equation:
\begin{equation}
\label{eq:30}
P_i\left(t+\Delta t\right)=P_i\left(t\right) \left[1-\sum_{j \neq i}
  \Pi_{i \to j} ^{\Delta t} \right]+  
\sum_{j\neq i} P_j\left(t\right)  \Pi_{j\to i}^{\Delta t}~,
\end{equation}
where $\Pi_{i \to j} ^{\Delta t}$ is the propagator of the random
walk, defined as the probability that the walker moves from vertex $i$
to vertex $j$ in a time interval $\Delta t$.  At any time $t$, node
$i$ will be linked to node $j$ if node $i$ becomes active and chooses
to connect to node $j$ (with probability $ma_i\Delta t/N$) or if node
$j$ becomes active and connects to node $i$ (with probability
$ma_j\Delta t/N$). In the first case, the instantaneous average degree
of node $i$, conditioned to the fact that it has become active, is
$k_{i}=m\left(1+\langle a\rangle\Delta t\right)$, while in the second
case we have $k_{i}=1+m\langle a\rangle\Delta t$, where the average is
conditioned to the fact that a vertex $j$ has fired and has connected
to $i$. A walker in node $i$ will then have to chose which one of the
$k_{i}$ connections to follow. We focus on homogenous random walks. In this case, the probability of moving from node
$i$ to node $j$ is inversely proportional to $k_{i}$. Thus the
propagator can be written as
\begin{eqnarray}
  \Pi_{i \to j} ^{\Delta t} &=& \frac{m a_i \Delta
    t}{N}\frac{1}{m\left(1+\langle a\rangle\Delta t\right)} 
  + \frac{m a_j \Delta t}{N} \frac{1}{1+m\langle a\rangle\Delta t}
  \nonumber \\
  &\simeq& \frac{\Delta t}{N}\left(a_i+ma_j\right),
  \label{eq:28}
\end{eqnarray}
where in the last expression we have neglected terms of order higher
than $\Delta t$.

In order to obtain a system level description it is convenient to
group nodes in the same activity class $a$, assuming that they are statistically equivalent, i.e.  considering the limit $N \to \infty$~\cite{barrat08-1,alex12-1}. Let
us define the number of walkers in a given node of class $a$ at time
$t$ as $W_a\left(t\right)=\left[NF\left(a\right)\right]^{-1}W\sum_{i
  \in a}P_i\left(t\right)$, where $W$ is the total number of walkers
in the systems. Considering Eq.~(\ref{eq:30}) in the limit $\Delta t
\rightarrow 0$ we can write:
\begin{eqnarray}
  \label{eq:7}
  \frac{\partial W_a\left(t\right)}{\partial t}&=&-a W_a\left(t\right)
  + a m w \nonumber\\ 
  &&- m \langle a\rangle W_a\left(t\right) +  \int a'
  W_{a'}\left(t\right) F\left(a'\right)\mathrm{d} a', 
\end{eqnarray}
where $w \equiv W/N$ is the average density of walkers per node, and
we have considered the continuous $a$ limit. The first two terms are
contributions due the activity of the nodes in class $a$, active nodes
which release all the walkers they have and receive walkers
originating from all the others nodes. The final two terms represent
the contribution to inactive nodes due to the activity of the nodes in
all the other classes. The stationary state of the process is defined
by the infinite time limit
$\lim_{t\to\infty}\dot{W}_a\left(t\right)=0$.  Using this condition in
Eq.~(\ref{eq:7}) we find the stationary solution
\begin{equation}
\label{beauty}
W_a=\frac{am w +\phi}{a+m\langle a\rangle},
\end{equation}
characterizing the stationary state of the random walk process in
activity driven networks, where $\phi=\int aF\left(a\right)W_a
\mathrm{d}a$ is the average number of walkers moving out of active nodes.
In the stationary state this quantity is constant, and we
can evaluate it self-consistently, which implies the equation:
\begin{equation}
\label{phi}
\phi=\int aF\left(a\right)\frac{am w +\phi}{a+m\langle
  a\rangle}\mathrm{d}a. 
\end{equation}
By considering heavy-tailed activity
distributions of the form $F\left(a\right)\sim a^{-\gamma}$, the explicit solution for $\phi$ can be written in terms
of hypergeometric functions which can be numerically evaluated. Heavy-tailed activity distribution have 
been empirically measured in real-world time varying networks.~\cite{perra12-1}.

\begin{figure}
\begin{center}
\includegraphics*  [width=0.45\textwidth] {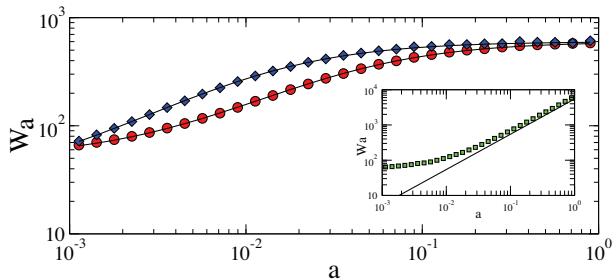}
\caption{Main plot: Stationary density $W_a$ of random walkers in
  activity driven networks with activity distribution $F(a) \sim
  a^{-\gamma}$. We consider $\gamma=2$ (circles) and $\gamma=2.8$
  (diamonds).  Solid lines represent the analytical prediction
  Eq.~(\ref{beauty}). Inset: Stationary density $W_a$ for random walks
  on top of an activity driven network with $F\left( a\right)\sim
  a^{-2} $, integrated over $T=50$ time steps. The solid line
  corresponds to the curve $W_a \sim a$, fitting the simulation points
  for large value of $a$.  Simulation parameters: $N=10^5$, $m=6$,
  $\epsilon=10^{-3}$ and $w =10^2$. Averages performed over
  $10^3$ independent simulations. }
\label{wa}
\end{center}
\end{figure}

To support the results of the analytical treatment, we have performed
extensive Monte Carlo simulations of the random walk process in
activity driven networks with $N=10^5$ nodes, $m=6$, and $ w =10^2$ walkers. In particular, we consider a power law
distribution $F\left(a\right)\sim a^{-\gamma}$, with activity
restricted in the interval $a \in \left[\epsilon,1\right]$, to avoid
possible divergencies in the limit $a\to0$.  As shown in
Figure~\ref{wa}, the analytical solution reproduces with great accuracy the
simulation results.
It is worth noticing that in quenched and annealed networks the number
of walkers in each node of degree $k$, is a linear function of the
degree: $W_k \sim k$~\cite{noh04,barrat08-1}. However, in time varying
networks the number of walkers is not a linear function of the
activity, but saturates at large values of $a$.  The difference is due
to the properties of the instantaneous network, where the nodes with
high activity have on average $k\sim m$ connections at each time step,
and therefore a limited capacity for collecting walkers. This key
feature cannot be recovered from time-aggregated views of dynamical
networks. To clarify this question we compare numerical simulations of
walkers in a network obtained integrating the activity driven model
with $N=10^5$ nodes, $m=6$ and $F\left( a\right)\sim a^{-2}$ over
$T=50$ time steps with the results obtained in the instantaneous
network (see inset of Figure~\ref{wa}). The lack of a saturation is
simply an artifact of using the time-aggregated network and highlights
the importance of an appropriate consideration of the time-varying
feature of networks in the study of exploration and spreading
processes in dynamical complex networks.

\begin{figure}
  \begin{center}
    \includegraphics*[width=0.45\textwidth] {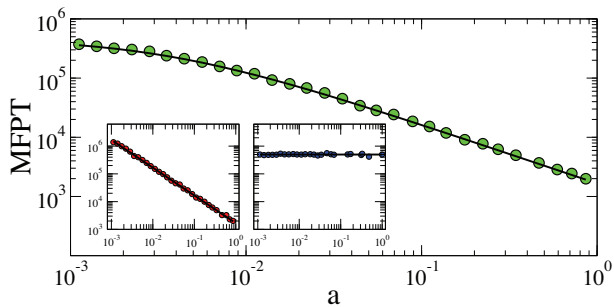}
    \caption{Main plot: MFPT of a random walker as a function of the activity $a$
      in activity driven networks with activity distribution $F\left(
        a\right)\sim a^{-2}$. Full line corresponds to the theoretical
      prediction Eq.~\eqref{mfpt_sol}. Right inset: MFPT as a function
      of activity for directed Type I activity driven networks. Left
      inset: MFPT as a function of activity for directed Type II
      activity driven networks.  Simulation parameters: $N=10^4$ ($N=10^3$ for Type I, blue dots),
      $m=6$, and $\epsilon=10^{-3}$.  Averages performed over $10^3$
      independent simulations for each activity class.}
\label{fig:mfpt}
\end{center}
\end{figure}

We focus next on the study the transport dynamics in such networks by
analyzing the mean first passage time (MFPT)~\cite{Redner01,noh04};
i.e. the the average time needed for a walker to arrive at node $i$
starting from any other node in the network. In other words, the
MFPT is the average number of steps needed to reach/find a specific
target with obvious consequences for network discovery processes. Let
us define $p\left(i,n\right)$ as the probability that the walker
reaches the target node $i$ for the first time at time $t=n \Delta
t$. Since each node is able, in principle, to connect directly to any
other node, this quantity is given simply by $p\left(i,n \right)=\xi_i
\left(1-\xi_i\right)^{n-1}$, where $\xi_i$ is the probability that the
random walker jumps to node $i$ in a time interval $\Delta t$. From
Eq.~\eqref{eq:28}, the probability that a walker in vertex $j$ jumps
to $i$ in a time $\Delta t$ is given by $\Pi_{j \rightarrow i}^{\Delta
  t}$. Thus we can write $\xi_i = \sum_j (W_j/W) \Pi_{j \rightarrow
  i}^{\Delta t}$, where we have replaced the probably that a single
random walker is at node $j$ at time $t$ by its steady state value
$W_j/W$.  The MFPT can thus be estimated as:
\begin{equation}
  \label{mfpt_sol}
  T_i=\sum_{n=0}^{\infty} \Delta t \; n p\left(i,n\right)=\frac{\Delta
    t}{\xi_i}
  =\frac{NW}{m a_i W+\sum_{j }a_j W_j}. 
\end{equation}
Interestingly, the MFPT each node $i$ is inversely proportional to its
activity plus a constant contribution from all the other nodes, in
clear contrast with what happens in quenched and annealed networks
where $\xi_i$ is equivalent to the stationary state of the random
walk, $\xi_i = W_i /W$. As before, the underlying cause of
this fundamental difference is the fact that in activity driven
networks the walker can be trapped in a node with low activity for
several time steps. The form of $\xi_i$ must then  consider
explicitly the dynamical connectivity patterns to account for the
resulting delays.  Figure~\ref{fig:mfpt} confirms these results with
extensive Monte Carlo  matching the analytical results presented in
Eq.~(\ref{mfpt_sol}).

The previous approach can be readily extended to the case of directed
networks.  By using the activity driven framework is possible to
define two types of time-varying directed networks. When a node $i$ is
active the $m$ generated links could be outgoing edges (Type I) or
ingoing edges (Type II). For both types of directed networks it is
possible to write down the diffusion propagator by following the same
approach used in the undirected case. In particular, it is possible to
show that if we define $W^{I}_a\left(t\right)$ and
$W^{II}_a\left(t\right)$ as the number of walkers in networks of Type
I and II, respectively, their stationary values read as
\begin{equation}
  W^{I}_a=\frac{w}{a}\frac{1}{\langle\frac{1}{a}\rangle};
  \;\ \;\ W^{II}_a=a w \frac{1}{\langle a\rangle}. 
\end{equation}
While we will report the full calculation elsewhere, this result can
be intuitively understood by considering that in Type I networks
active nodes create outgoing links. Walkers are thus more likely to
diffuse out of these nodes, meaning that the higher the activity, the
smaller the number of walkers in the nodes of that class.  In Type II
networks, on the other hand, active nodes create ingoing
links. Walkers are thus more likely to diffuse into high activity
nodes and the scaling of the stationary state is linear with the
activity. Interestingly the undirected functional form of the
stationary state is a combination of these two different behaviors.

By following the same reasoning used for the undirected case it is
straightforward to derive the analytic expression of the MFTP for directed
activity networks, namely:
\begin{equation}
\label{mfpt_dir}
T_i^{I}=\frac{NW}{\sum_{j} a_j W_j} ; \;\ \;\
T_i^{II}=\frac{N}{m a_i}. 
\end{equation}
In the first case the MFPT is independent of the activity of the
considered node. The walker can move to node $i$ just when other
active nodes create outgoing links pointing to $i$. In the second
case, the MFPT is just proportional to the activity of the node $i$
and is not a function of the activity of the others nodes. Also in this
case we recover that the propagator of the random walk for undirected
activity driven networks has these two symmetric contributions that
both contribute to the undirected MFPT.

The analytical results can be validated by means of Monte Carlo
simulations. The right  inset of Figure~\ref{fig:mfpt}
refers to activity networks of Type I. We fix $N=10^3$,  $m=6$, one
walker and activities distributed according to a power-law $F\left(
  a\right)\sim a^{-2}$. We then measured the MFPT selecting $10^3$
targets for each activity class. The simulations are indistinguishable
from the analytical result in Eq.~(\ref{mfpt_dir}). The left inset in 
Fig.~\ref{fig:mfpt} refers to activity networks of Type
II under the same simulation parameters except for the number of nodes fixed to $N=10^4$ in this case. Again a perfect match is
observed with the analytical result Eq.~(\ref{mfpt_dir}).

From the above results it is evident that the dynamics of time-varying
networks significantly alters the standard picture achieved for
dynamical processes in static networks. Focusing on the specific case
of activity driven networks and the simple random walk process, the
present results open the path to a number of future studies where the
dynamics of the network will have to be considered, in order to avoid
misleading results in the analysis of dynamical processes in most
situation of practical interest. Finally, we note that the time
varying network model we have considered is Markovian (memoryless) and
lacking dynamical correlations, rife in real dynamical networks
\cite{holme11-1}. The investigation of the effects of these relevant
properties on diffusion calls for additional research efforts.

\begin{acknowledgments}
  The work has been partly sponsored by the Army Research Laboratory
  and was accomplished under Cooperative Agreement Number
  W911NF-09-2-0053. R.P.-S. acknowledges financial support from the Spanish MICINN (FEDER),
  under project No. FIS2010-21781-C02-01, and 
  ICREA Academia, funded by the Generalitat de Catalunya.
\end{acknowledgments}


%

\end{document}